\def\eess{1} 
\def\conf{0} 
\def\noTitle{0} 
\def\latexversion{2} 
\ifnum\conf=1
\else
\fi 

\ifnum\latexversion=1
\documentstyle[11pt,fullpage]{article}
\input{psfig} 
\else 
\documentclass[11pt]{article}
\usepackage{amsfonts,fullpage,epsfig,color}
\fi 

\title{{\large\bf 
Testing Bipartitness in an Augmented VDF Bounded-Degree Graph Model}%
\thanks{Preliminary version; comments are most welcome.}}
\author{Oded Goldreich\thanks{Faculty of Mathematics and Computer Science,  
   Weizmann Institute of Science, Rehovot, {\sc Israel}.
   Email: {\tt oded.goldreich@weizmann.ac.il}.}}

\ifnum\conf=0
\else 
\fi

\newtheorem{theo}{Theorem}[section] 
\newtheorem{prop}[theo]{Proposition}
\newtheorem{rem}[theo]{Remark}
\newtheorem{lem}[theo]{Lemma}
\newtheorem{define}[theo]{Definition}
\newtheorem{conj}[theo]{Conjecture}        
\newtheorem{coro}[theo]{Corollary}
\newtheorem{open}[theo]{Open Problem}
\newtheorem{obs}[theo]{Observation}
\newtheorem{alg}[theo]{Algorithm}
\newtheorem{claim}[theo]{Claim}
\newtheorem{techdef}{Definition}[theo] 
\newtheorem{techclm}[techdef]{Claim}
\newtheorem{fact}[techdef]{Fact}
\newtheorem{techalg}[techdef]{Algorithm}        

\newcommand{\BE}{\begin{enumerate}} \newcommand{\EE}{\end{enumerate}}
\newcommand{\BI}{\begin{itemize}} \newcommand{\EI}{\end{itemize}}
\newcommand{\BDes}{\begin{description}}\newcommand{\EDes}{\end{description}}
\newcommand{\BT}{\begin{theo}} \newcommand{\ET}{\end{theo}}
\newcommand{\BL}{\begin{lem}} \newcommand{\EL}{\end{lem}}
\newcommand{\BD}{\begin{define}} \newcommand{\ED}{\end{define}}
\newcommand{\BCJ}{\begin{conj}} \newcommand{\ECJ}{\end{conj}}
\newcommand{\BCO}{\begin{coro}} \newcommand{\ECO}{\end{coro}}
\newcommand{\BOP}{\begin{open}} \newcommand{\EOP}{\end{open}}
\newcommand{\BO}{\begin{obs}} \newcommand{\EO}{\end{obs}}
\newcommand{\BR}{\begin{rem}} \newcommand{\ER}{\end{rem}}
\newcommand{\BCM}{\begin{claim}} \newcommand{\ECM}{\end{claim}}
\newcommand{\BA}{\begin{alg}} \newcommand{\EA}{\end{alg}}
\newcommand{\Balg}{\begin{techalg}} \newcommand{\Ealg}{\end{techalg}}
\newcommand{\Bcm}{\begin{techclm}} \newcommand{\Ecm}{\end{techclm}}
\newcommand{\Bf}{\begin{fact}} \newcommand{\Ef}{\end{fact}}
\newcommand{\Bdf}{\begin{techdef}} \newcommand{\Edf}{\end{techdef}}
\newcommand{\BP}{\begin{prop}} \newcommand{\EP}{\end{prop}}
\ifnum\conf=0
\newtheorem{xtechdef}[techdef]{Definition}
\newtheorem{xtechclm}[techdef]{Claim}
\else
\newtheorem{xtechdef}[theo]{Definition}
\newtheorem{xtechclm}[theo]{Claim}
\fi
\newcommand{\xBCM}{\begin{xtechclm}}
\newcommand{\xECM}{\end{xtechclm}}
\newcommand{\xBDF}{\begin{xtechdef}}
\newcommand{\xEDF}{\end{xtechdef}}

\def\FullBox{\hbox{\vrule width 8pt height 8pt depth 0pt}}
\def\sFullBox{\hbox{\vrule width 6pt height 6pt depth 0pt}}
\newcommand{\qed}{\;\;\;\FullBox}
\ifnum\latexversion=1
\newcommand{\qqed}{\(\;\;\;\Box\)}
\else
\newcommand{\qqed}{\(\;\;\;\sFullBox\)}
\fi 
\newenvironment{proof}{\noindent{\bf Proof:~~}}{\(\qed\)}
\newenvironment{proofsk}{\noindent{\bf Proof Sketch:~}}{\(\qed\)}
\newcommand{\BPF}{\begin{proof}} \newcommand{\EPF}{\end{proof}}
\newcommand{\BPFS}{\begin{proofsk}} \newcommand{\EPFS}{\end{proofsk}}
\newenvironment{proofof}[1]{\noindent{\bf Proof of {#1}:~~}}{\(\qed\)}
\ifnum\conf=0
\newcommand{\xBPF}{\begin{proof}} \newcommand{\xEPF}{\end{proof}}
\else
\newcommand{\xBPF}{\BPFS} \newcommand{\xEPF}{\EPFS} 
\fi
\newcommand{\BPFOF}{\begin{proofof}} \newcommand {\EPFOF}{\end{proofof}}

\newenvironment{techproof}{\noindent{\sf Proof:~~}}{\qqed}
\newcommand{\Bpf}{\begin{techproof}} \newcommand{\Epf}{\end{techproof}}

\newcommand{\BEQ}{\begin{equation}} \newcommand{\EEQ}{\end{equation}}
\newcommand{\BEQN}{\begin{eqnarray}}\newcommand{\EEQN}{\end{eqnarray}}

\newlength{\saveparindent}
\newlength{\saveparskip}
\ifnum\latexversion=1
  \font\tenmsb=msbm10 scaled\magstep1
  \font\sevenmsb=msbm7 scaled\magstep1
  \font\fivemsb=msbm5 scaled\magstep1
  \newfam\msbfam
  \textfont\msbfam=\tenmsb
  \scriptfont\msbfam=\sevenmsb
  \scriptscriptfont\msbfam=\fivemsb
  
  \newcommand{\N}{\Bbb N}
\else
\newcommand{\N}{{\mathbb{N}}}
\fi

\newcommand{\D}{{\cal D}}
\newcommand{\sD}{{{\tt samp}_{\cal D}}}
\newcommand{\eD}{{{\tt eval}_{\cal D}}}

\newcommand{\poly}{{\rm poly}}

\newcommand{\prob}{{\rm Pr}}

\newcommand{\Exp}{{\rm E}}

\newcommand{\eqdef}{\stackrel{\rm def}{=}}

\newcommand{\eqref}[1]{Eq.~{\rm(\ref{#1})}}
\newcommand{\ceil}[1]{{\lceil{#1}\rceil}}
\newcommand{\floor}[1]{{\lfloor{#1}\rfloor}}
\newcommand{\round}[1]{{\lfloor{#1}\rceil}}
\newcommand{\ang}[1]{{\langle{#1}\rangle}}

\newcommand{\e}{\epsilon}

\newcommand{\bitset}{{\{0,1\}}}

\renewcommand{\Exp}{{\rm Exp}}

\newcommand{\tildeO}{{\widetilde{O}}}

\newcommand{\nn}{{\widetilde{n}}}

\newcommand{\xth}{{\rm th}}

\begin{document}

\ifnum\noTitle=1 
\begin{flushright}Oded: \today \end{flushright}
\else 

\begin{titlepage}
\maketitle

\begin{abstract}
In a recent work ({\em ECCC}, TR18-171, 2018), 
we introduced models of testing graph properties in which,
in addition to answers to the usual graph-queries, 
the tester obtains 
{\em random vertices drawn according to an arbitrary distribution $\D$}. 
Such a tester is required to distinguish between graphs that 
have the property and graphs that are far from having the property,
{\em where the distance between graphs is defined based on the
unknown vertex distribution $\D$}. 
These (``vertex-distribution free'' (VDF)) models 
generalize the standard models 
in which $\D$ is postulated to be uniform on the vertex-set,
and they were studies both in the dense graph model
and in the bounded-degree graph model. 

The focus of the aforementioned work was on testers, called {\sf strong},
whose query complexity depends only on the proximity parameter $\e$.
Unfortunately, in the standard bounded-degree graph model,
some natural properties such as Bipartiteness do not have strong testers,   
and others (like cycle-freeness) do not have strong testers
of one-sided error (whereas one-sided error was shown inherent 
to the VDF model). 
Hence, it was suggested to study general (i.e., non-strong) testers 
of ``sub-linear'' complexity. 

In this work, we pursue the foregoing suggestion,
but do so in a model that augments the model presented in
the aforementioned work. 
Specifically, we provide the tester with an evaluation oracle
to the unknown distribution $\D$, in addition to samples of $\D$
and oracle access to the tested graph. 
Our main results are testers for Bipartitness and cycle-freeness, 
in this augmented model, having complexity that is almost-linear
in the square root of the ``effective support size'' of $\D$. 
\end{abstract}

\pagestyle{empty}
\vfill


\vfill 
\footnotesize\tableofcontents\normalsize
\vfill 

\end{titlepage}


\pagenumbering{arabic}

\fi

\section{Introduction} 
In the last couple of decades, 
the area of property testing has attracted much attention 
(see, e.g., a recent textbook~\cite{G:pt}). 
Loosely speaking, property testing typically refers to sub-linear time
probabilistic algorithms for deciding whether a given object has 
a predetermined property or is far from any object having this property. 
Such algorithms, called testers, obtain local views of the object
by making adequate queries; that is, the object is seen as a function 
and the testers get oracle access to this function
(and thus may be expected to work in time that 
is sub-linear in the size of the object). 

A significant portion of the foregoing research was devoted
to testing graph properties in two different models:
the dense graph model 
(introduced in~\cite{GGR} and reviewed in~\cite[Chap.~8]{G:pt})  
and the bounded-degree graph model 
(introduced in~\cite{GR} and reviewed in~\cite[Chap.~9]{G:pt}). 
In both models, it was postulated that the tester can sample
the vertex-set uniformly at random%
\footnote{Actually, in all these models, it is postulated
that the vertex-set consists of $[n]=\{1,2,...,n\}$,
where $n$ is a natural number that is given explicitly to the tester,
enabling it to sample $[n]$ uniformly at random.}
(and, in both models, distances between graphs 
were defined with respect to this distribution). 

In a recent work~\cite{G:vdf}, 
we considered settings in which uniformly sampling the vertex-set
of the graph is not realistic, and asked what happens if the tester
can obtain random vertices drawn according to some distribution $\D$
(and, in addition, obtain answers to the usual graph-queries). 
The distribution $\D$ should be thought of as arising from 
some application, and it is not known {\em a priori}\/
to the (application-independent) tester.
In this case, it is also reasonable to define the distance 
between graphs with respect to the distribution $\D$,
since this is the distribution that the application uses. 

These considerations led us to introduce models of testing graph properties 
in which the tester obtains {\em random vertices drawn according to 
an arbitrary vertex distribution $\D$}\/ 
(and, in addition, obtains answers to the usual graph-queries). 
Such a tester is required to distinguish between graphs that 
have the property and graphs that are far from having the property,
{\em where the distance between graphs is defined based on the
unknown vertex distribution $\D$}. 
These (``vertex-distribution free'' (VDF)) models 
generalize the standard models 
in which $\D$ is postulated to be uniform on the vertex-set,
and they were studies both in the dense graph model
and in the bounded-degree graph model
(see~\cite[Sec.~2]{G:vdf} and~\cite[Sec.~3]{G:vdf}, respectively).  

The focus of~\cite{G:vdf} was on testers, called {\sf strong},
whose query complexity depends only on the proximity parameter $\e$.
Unfortunately, in the standard {\em bounded-degree graph model},
some natural properties such as bipartiteness do not have strong testers,   
and others (like cycle-freeness) do not have strong testers
of one-sided error (whereas one-sided error was shown inherent 
to the VDF model~\cite[Thm.~1.1]{G:vdf}).
Hence, it was suggested in~\cite[Sec.~5.2]{G:vdf} 
to study general (i.e., non-strong) testers 
of ``sub-linear'' complexity,
especially for the VDF bounded-degree graph model.  

In this work, we pursue the foregoing suggestion,
but do so in a model that augments the model presented in~\cite{G:vdf}. 
Specifically, we provide the tester with an evaluation oracle
to the unknown distribution $\D$, in addition to samples of $\D$
and oracle access to the tested graph. 

\subsection{The vertex-distribution-free model and its augmentation} 
\label{vdf-model:sec} 
We start by recalling the vertex-distribution free (VDF) model 
that generalizes the bounded-degree graph model. 
Essentially, this model differs from the standard 
bounded-degree graph model in that the tester cannot obtain
uniformly distributed vertices, but rather 
random vertices drawn according to an arbitrary distribution $\D$
(which is unknown {\em a priori})). 
(In addition, the tester obtains answers to the usual graph-queries.) 
As usual, the tester is required to accept (whp) graphs that
have the predetermined property and reject (whp) graph that
that are far from the property, but the distance between graphs
is defined in terms of the distribution $\D$.
Specifically, when generalizing the standard bounded-degree model,
we define the distance between graphs as the sum of the
weights of the edges in their symmetric difference,
where {\em the weight of an edge is proportional to the sum of
the probability weights of its end-points according to $\D$}. 

Recall that the bounded-degree model 
(both in its standard and VDF incarnations)
refers to a fixed degree bound, denoted $d$, 
and to graphs that are represented by their incidence functions;
that is, the graph $G=(V,E)$ is represented by the incidence
function $g:V\times[d]\to V\cup\{\bot\}$, 
where $g(v,i)=u$ if $u$ is the $i^\xth$ neighbour of $v$
and $g(v,i)=\bot$ if $v$ has less than $i$ neighbours.
Fixing a vertex distribution $\D$, 
we say that the graph $G$ or rather its incidence function $g$
is {\sf $\e$-far} from the graph property $\Pi$ if,
for every $g':V\times[d]\to V\cup\{\bot\}$ that represents a graph in $\Pi$,  
it holds that $\prob_{v\gets\D,i\in[d]}[g(v,i)\!\neq\!g'(v,i)]>\e$.

Hence, following~\cite[Def.~3.1]{G:vdf}, 
a {\sf tester of $\Pi$ in the VDF bounded-degree graph model} 
is given a proximity parameter $\e$, 
samples drawn from an arbitrary distribution $\D$,
and oracle access to the incidence function of the graph, $G=(V,E)$.
It is required that, for every vertex-distribution $\D$
(and every $\e>0$ and $G$), 
the tester accepts (whp) if $G$ is in $\Pi$
and rejects (whp) if $G$ is $\e$-far from $\Pi$
(where the distance is defined according to $\D$). 

\paragraph{The augmentation.} 
Here, we augment the foregoing model by providing the tester 
also with an {\em evaluation oracle}\/ to the vertex distribution; 
that is, an oracle that on query $v$ 
returns $\D(v)=\prob_{x\gets\D}[x\!=\!v]$.
This augmentation is introduced because we could not obtain
our results without it (or without some relaxation of it),
but it can be justified as feasible in some settings
(see brief discussion in Section~\ref{discuss:sec}). 
At this point, we spell out the resulting definition of a tester.

\BD{\em(the augmented VDF testing model):}
\label{aug-vdf-model:def}
For a fixed $d\in\N$, let $\Pi$ be a property of graphs 
of degree at most $d$. 
An {\sf augmented VDF tester for the graph property $\Pi$ 
(in the bounded-degree graph model)} 
is a probabilistic oracle machine $T$ 
that satisfies the following two conditions 
{\rm(for all sufficiently large $V$)},
when given access to the following three oracles: 
an incidence function $g:V\times[d]\to V\cup\{\bot\}$,  
a device~-- denoted $\sD$~-- that samples in $V$
according to an arbitrary distribution $\D$, 
and an evaluation oracle~-- denoted $\eD$~-- for $\D$.   
\BE
\item
The tester accepts each $G=(V,E)\in\Pi$ with probability at least $2/3$; 
that is, for every $g:V\times[d]\to V\cup\{\bot\}$ representing 
a graph in $\Pi$ and every $\D$ {\em(and $\e>0$)},
it holds that $\prob[T^{g,\sD,\eD}(\e)\!=\!1]\geq2/3$.
\item
Given $\e>0$ and oracle access to any $G=(V,E)$ and $\D$
such that $G$ is $\e$-far from $\Pi$ according to $\D$,
the tester rejects with probability at least $2/3$; 
that is, for every $\e>0$ and distribution $\D$,
if $g:V\times[d]\to V\cup\{\bot\}$ satisfies $\delta_\D^\Pi(g)>\e$,
then it holds that $\prob[T^{g,\sD,\eD}(\e)\!=\!0]\geq2/3$, 
where $\delta_\D^\Pi(g)$ denotes the minimum of $\delta_\D(g,g')$
taken over all incidence functions $g':V\times[d]\to V\cup\{\bot\}$
that represent graphs in $\Pi$, and 
\BEQ\label{dist:eqdef}
\delta_\D(g,g')\eqdef\prob_{v\gets\D,i\in[d]}[g(v,i)\neq g'(v,i)]. 
\EEQ
{\rm(That is,
$\delta_\D(g,g')
 =\sum_{v\in V}\D(v)\cdot|\{i\in[d]:g(v,i)\neq g'(v,i)\}|/d$.)} 
\EE
The tester is said to have {\sf one-sided error probability}
if it always accepts graphs in $\Pi$; 
that is, for every $g:V\times[d]\to V\cup\{\bot\}$ representing 
a graph in $\Pi$ {\em(and every $\D$ and $\e>0$)},
it holds that $\prob[T^{g,\sD,\eD}(\e)\!=\!1]=1$.
\ED
At times, we shall identify the incidence function $g$ 
with the graph $G$ that $g$ represents,
and simply say that we provide the tester with oracle access to $G$. 

The {\sf query complexity} of a tester is the maximum number 
of queries it makes to its (three) oracles as a function of $\e$ 
and parameters of the vertex-distribution $\D:V\to[0,1]$.
The parameters we have in mind are label-invariant,
where a parameter $\psi$ is {\sf label invariable}
if $\psi(\D)=\psi(\D')$ for any two distributions $\D$ and $\D'$ 
that have the same histogram (i.e., for every $p>0$ it holds 
that $|\{v:\D(v)\!=\!p\}|=|\{v:\D'(v)\!=\!p\}|$).%
\footnote{Indeed, in this case there exists a permutation $\pi:V\to V$
such that $\D'(v)=\D(\pi(v))$ for every $v\in V$.}
Recall that in the standard testing model, 
the complexity could depend on the size of the vertex-set, 
which is a special case of a parameter of $\D:V\to[0.1]$. 
However, we are interested in more refined parameters of $\D$.
The first parameter that comes to mind is the support size of $\D$,
but this parameter is too sensitive to insignificant changes in $\D$
(e.g., any distribution over $V$ is infinitesimally close
to having support size $|V|$). 
A more robust parameter is the ``minimum effective support size'';
that is, being ``close'' to a distribution 
with the specified support-size (cf.,~\cite{BCG}). 

\BD{\em(effective support size):}
\label{eff-support:def}
We say that the distribution $\D$ 
has {\sf $\eta$-effective support of size $n$}
if $\D$ is $\eta$-close to a distribution 
that has support size at most $n$,
where $\D$ is {\sf $\e$-close} to $\D'$ 
if their total variation distance is at most $\e$.
The {\sf minimal $\eta$-effective support size of $\D$} is the 
minimal $n$ such that $\D$ has $\eta$-effective support of size $n$.
\ED
The notion of effective support size is much more robust 
that the support size; in particular, if $\D$ is infinitesimally close
to a distribution that has $\eta$-effective support of size $n$,
then $\D$ that has $\eta$-effective support of size $n+1$
(where the additional unit is needed only in pathological cases).%
\footnote{Let $\D'$ be the foregoing distribution 
that has $\eta$-effective support of size $n$. 
Then, the typical case is that, for some $\eta'<\eta$, 
the distribution $\D'$ has $\eta'$-effective support of size $n$. 
In this case, any distribution that is $(\eta-\eta')$-close to $\D'$
has $\eta$-effective support of size $n$. 
The pathological case is that $\D'$ has $\eta$-effective support of size $n$,
but for every $\eta'<\eta$ the minimal $\eta'$-effective support size of $\D'$
is larger than $n$. We claim that in this case, for some $\eta'<\eta$,
the distribution $\D'$ has $\eta'$-effective support of size $n+1$
(and it follows that any distribution that is $(\eta-\eta')$-close to $\D'$
has $\eta$-effective support of size $n+1$). 
Suppose that $\D'$ is $\eta$-close to a distribution $\D''$ 
of support size $n$. We prove the claim by considering two cases. 
\BE
\item If the support of $\D'$ is contained in the support of $\D''$, 
then the claim is trivial (since then $\D'$ has support size $n$).
\item Otherwise, let $v$ be in the support of $\D'$ but not
in the support of $\D''$, and consider modifying $\D''$ by moving 
a probability mass of $\D'(v)>0$ from $\{u:\D''(u)>\D'(u)\}$ to $v$. 
Then, the modified distribution $\D'''$ has support size $n+1$
and is $(\eta-\D'(v))$-close to $\D'$, and so the claim follows
with $\eta'=\eta-\D'(v)$. 
\EE}
In general, if $\D$ is $o(\e)$-close 
to a distribution that has $\e$-effective support size $n$,
then $\D$ that has $(1+o(1))\cdot\e$-effective support size $n$. 


\paragraph{An initial observation and an open problem.} 
Recall that it was shown in~\cite[Prop.~3.2]{G:vdf} that,
without loss of generality, any tester in the VDF model
only queries vertices that were provided as answers to prior
sample and graph queries. 
The argument extends to the augmented VDF model.
In contrast, it is unclear whether~\cite[Thm.~3.3]{G:vdf},
which asserts that strong testability in the VDF model
yields strong testability with one-sided error,
holds in the augmented VDF model.
 
\BOP{\em(does one-sided error testing reduce to general testing):} 
\label{ose:open}
For $q:(0,1]\to\N$, suppose that $\Pi$ is a graph property that 
can be tested using $q(\e)$ queries in the augmented VDF model,
where $\e$ denotes the proximity parameter. 
Does there exists a function $q':(0,1]\to\N$ such that $\Pi$ 
has a one-sided error tester of query complexity $q'(\e)$ 
in the augmented VDF model.
\EOP 
(Recall that in the original VDF model,
an upper bound of $q'(\e)=\exp(O(q(\e)))$ was shown~\cite[Thm.~3.3]{G:vdf}.)

\subsection{Our results}
Our main result is testing Bipartitness in the augmented VDF model
within complexity that matches the complexity of the known tester
in the standard (bounded-degree graph) model~\cite{GR2},
which in turn is almost optimal~\cite{GR}.

\BT{\em(testing bipartitness in the augmented VDF model):}
\label{bipartite:thm}
Bipartiteness can be tested in the augmented VDF testing model
{\rm(of Definition~\ref{aug-vdf-model:def})} 
in expected time $\tildeO({\sqrt n})\cdot\poly(1/\e)$,
where $n$ denotes the minimal $\e/5$-effective support size
of the vertex distribution $\D$, used by the tester. 
Furthermore, the tester has one-sided error. 
\ET
The tester that we use when proving Theorem~\ref{bipartite:thm} 
starts by obtaining a rough approximation 
of the minimal effective support size of $\D$.
In fact, a rough approximation of the effective support size
of $\D$ is implicit in the running-time of the asserted tester.  
We comment that it is essential to make queries to $\eD$ in order 
to obtain such an approximation, since obtaining such an approximation 
by making queries only to $\sD$ has complexity $n^{1-o(1)}$ 
(cf.~\cite{RRSS}).\footnote{Specifically,
at least $n^{1-o(1)}$ queries are necessary to distinguish 
an $n$-grained distribution of support size $n/11$ from 
an $n$-grained distribution with support size $n/f$, 
for any $f=n^{o(1)}$, where a distribution is called $n$-grained
if all probabilities are multiples of $1/n$.}
In fact, using both types of queries,
we obtain a good approximation in polylogarithmic time 
(see Section~\ref{approx-eff-support:sec}). 

Next, we extend the known reduction of testing cycle-freeness
to testing Bipartiteness, presented in~\cite{CGRSSS}
for the standard (bounded degree graph) model,
to the (augmented) VDF model.
Combining this reduction with Theorem~\ref{bipartite:thm}, we obtain

\BT{\em(testing cycle-freeness in the augmented VDF model):}
\label{cycle-free:thm}
Cycle-freeness can be tested in the augmented VDF testing model
{\rm(of Definition~\ref{aug-vdf-model:def})} 
in expected time $\tildeO({\sqrt n})\cdot\poly(1/\e)$,
where $n$ denotes the minimal $\e/5$-effective support size
of the vertex distribution $\D$ used by the tester. 
Furthermore, the tester has one-sided error. 
\ET
A begging open problem is whether the foregoing results
can also be obtained in the original VDF model (of~\cite{G:vdf}),
at least when providing the tester with the effective support size.

\BOP{\rm(the complexity of testing Bipartiteness in the VDF model):} 
\label{bipartite:open}
What is the query complexity of testing Bipartiteness 
in the {\rm(original)} VDF model.  
%
In particular, can Bipartite be tested in this model 
in time $\tildeO({\sqrt n})\cdot\poly(1/\e)$,
where $n$ denotes the $\Omega(\e)$-effective support size
of the vertex distribution $\D$ used by the tester? 
The question holds both for the VDF model
{\rm(as defined in~\cite{G:vdf})},
and in a model in which the tester is provided 
with the effective support size.%
\footnote{Note that the result of~\cite{RRSS} does not rule out 
the possibility of approximating $n$ to a factor of $n^c$
using $n^{0.5+c'}$ queries, for some $c,c'\in(0,0.5)$.}
\EOP 
Since the reduction of testing cycle-freeness to Bipartiteness
does not make queries to $\D$ (and is in fact oblivious of $\D$), 
any upper bound regarding testing Bipartiteness
in the (original) VDF model would yield a similar result for
testing cycle-freeness. However, the latter problem may be easier. 

\subsection{Techniques}\label{tech:sec} 
The testers asserted in Theorems~\ref{bipartite:thm} 
and~\ref{cycle-free:thm} constitute testers for the standard 
(bounded-degree graph) model that meet the best results known 
in that model. Given that the proofs of the latter results
are quite complex (see, e.g.,~\cite{GR2}), 
it is fortunate that we can proceed by reducing 
the current results to the known ones. 

\paragraph{Proving Theorem \protect\ref{bipartite:thm}.} 
The Bipartite tester asserted in Theorem~\ref{bipartite:thm} 
is obtained by a natural adaptation of the corresponding tester
for the standard (bounded-degree graph) model~\cite{GR2}. 
The latter tester operates by taking many short random walks from
few randomly selected vertices, where, in each step of a random walk, 
the next vertex is selected uniformly among the neighbors
of the current vertex. Instead, our tester will select the 
the next vertex (among the neighbors of the current vertex)
with probability that is {\em proportional to the probability weight 
of the corresponding incident edges}\/ (according to $\D$);
that is, being at vertex $v$ we move to a neighbor $w$
with probability proportional to $\D(v)+\D(w)$.
Here is where we make use of the evaluation oracle $\eD$. 
(The start vertices will be selected with probability that
is proportional to the $\D$-weight of their incident edges.) 

Since the analysis of the foregoing tester in the standard
(bounded-degree graph) model is quite complex, 
we wish to use this analysis (of~\cite{GR2}) as a black-box. 
Towards this end, 
we view the foregoing tester as emulating the tester of~\cite{GR2}
on an auxiliary graph in which weighted edges are replaced
by a proportional number of parallel edges,
while recalling that the analysis of~\cite{GR2} holds also 
for (non-simple) graphs having parallel edges (see~\cite{KKR}). 
(We stress that this is a mental experiment performed in the analysis;
the actual algorithm is essentially as outlined above.) 

Indeed, this strategy may yield an auxiliary graph in which 
the average degree is much smaller than the maximum degree,
whereas the analysis in~\cite{GR2} relies on the two quantities
being of the same order of magnitude (see~\cite{KKR}). 
This issue can be addressed by using the transformation of~\cite{KKR},
but doing so will yield a slightly less natural tester
(than the one outlined above).
So instead of applying the transformation of~\cite{KKR}
as a black-box, we use an alternative implementation of 
its underlying ideas, which capitalizes on the fact
that we can set the number of parallel edges 
(in our mental experiment) to be much larger
than the number of vertices in the tested graph.
We stress that the complexity of the tester of~\cite{GR2}
is dominated by the number of the vertices in the graph,
and hardly depends on the number of edges in it.%
\footnote{The complexity is almost-linear in a square root 
of the number of vertices, 
and only grows logarithmically with the number of edges.}

The forgoing description presumes that we have 
a good upper bound on the support size of $\D$. 
(Note that such an upper bound (along with oracle access to the 
auxiliary graph) suffices for emulating the tester of~\cite{GR2}.)
Actually, aiming at complexity bounds that depend on 
the effective support size of $\D$ rather than on its support size, 
we ``trim'' the graph by ignoring edges of weight $o(\e/n)$,
where $n$ is an upper bound on the $\e/4$-effective support size of $\D$.
Lastly, we show how to obtain such a good upper bound
by using both the sampling and evaluation oracles of $\D$. 

As for approximating the effective support size of $\D$, 
the basic observation is that if $n$ is the minimal integer
such that an $\eta$ fraction of the weight of $\D$ resides 
on elements of weight at most $\eta/n$,
then the $\Theta(\eta)$-support size of $\D$
is between $\Omega(n)$ and $O(n/\eta)$. 
Such a rough approximation suffices for the foregoing application,
and we obtain it obtained by using a doubling procedure.
We actually obtain a better approximation 
(see Theorem~\ref{approx-eff:thm}) 
by using the rough approximation as a starting point 
and approximating the total weight assigned to each set 
$\{v:2^{-i+1}\leq\D(v)<2^{-i}\}$ for $i\in[\log(n/\eta)+O(1)]$. 

\paragraph{Proving Theorem \protect\ref{cycle-free:thm}.} 
The cycle-freeness tester asserted in Theorem~\ref{cycle-free:thm} 
is obtained by using the known reduction of testing cycle-freeness
to testing Bipartiteness, which was presented and analyzed
for the standard (bounded degree graph) model in~\cite{CGRSSS}. 
Recall that this reduction replaces each edge at random
either by a path of length two or by a path of length one
(equiv., leaves the edge intact). 
It was shown in~\cite[Lem~3.1]{CGRSSS} that this transformation
translates a graph that is far from being cycle-free to a graph
that is far from being bipartite, where the distance refers to the number 
of edges that should be omitted to make the graph satisfy the property.
The challenge is to extend this claim to weighted graphs
(or rather to distances as measured by the sum of the weights of 
edges that should be omitted to make the graph satisfy the property).

We meet this challenge by observing that the edges that should be omitted
in order to make the original graph cycle-free are those that
do not reside on a {\em maximal spanning forest of the graph}. 
We then bucket these edges according to their approximate weight, 
and consider only the edges that are in the union of 
the heaviest bucket and said spanning forest.
Considering the two-connected components of the corresponding graph,
we observe that the remaining edges of the forest 
(i.e., those that reside inside these two-connected components)
cannot be lighter than the edges in the bucket, 
which reduces the problem of lower-bounding the weight of edges 
to lower-bounding their number. 
At this point, we invoke~\cite[Lem~3.1]{CGRSSS} and are done. 

\subsection{Discussion}\label{discuss:sec} 
As admitted upfront, the augmentation of the VDF model
captured by Definition~\ref{aug-vdf-model:def} is made
for opportunistic reasons: 
This augmentation (or some relaxation of it)
seems essential to the testers asserted in
Theorems~\ref{bipartite:thm} and~\ref{cycle-free:thm}. 
Nevertheless, one may envision setting in which an evaluation
oracle as postulated in the augmentation can be implemented
or at least be well-approximated.

Recall that the VDF model was motivated by settings in which 
some process (or application) of interest refers to 
(or embeds or emulates) a huge graph; in particular, the process  
generates random vertices according to some unknown distribution $\D$,
and answers incidence queries regarding the graph. 
Furthermore, in case the vertices of the graph are real sites, 
they may maintain a count of the number of times they were visited
by the foregoing process (or application). 
This yields a good approximation of the visiting probabilities
that underly the vertex distribution $\D$ in question. 

Indeed, the vertex distribution represents the ``importance''
of the various vertices from the application's point of view;
that is, the application encounters vertices according to
the distribution $\D$, and the relative ``importance'' of 
a vertex (to the application) is captured by the probability 
that it is encountered (by the application). 
Hence, the distance of a graph to the property represents the relative 
importance of the ``part of the graph'' that violates the property. 

A VDF tester offers an application-independent way of determining 
whether the huge graph (embedded or emulated by an application) 
has some predetermined property or is far from having the property, 
where the distance that is relevant here is one that is induced by 
the vertex-distribution $\D$ (arising from the application).

\section{The Bipartiteness Tester}\label{bip:sec}
For starters, we consider a model in which the tester
is further augmented by providing it with the effective 
support size of the vertex distribution $\D$.
Specifically, on input proximity parameter $\e$,
we also provide the tester with an upper bound, denoted $n$,
on the minimal $\e/4$-effective support size of $\D$. 
The complexity of the following tester depends on that upper bound,
and at a latter stage we shall show how the tester can obtain 
a good upper bound by itself. 

\BT{\em(a tester that gets an upper bound on the effective support size):} 
\label{bip-core:thm}
There exists an oracle machine $T$ that, on input $\e$ and $n$
such that $n$ is an upper bound on the minimal $\e/4$-effective
support size of $\D$, and oracle access to $G,\sD,\eD$,
runs in time $\tildeO({\sqrt n})\cdot\poly(1/\e)$
and constitutes a bipartite tester {\em(of one-sided error)} 
as defined above. That is:  
\BE
\item 
If $G$ is bipartite, then $\prob[T^{G,\sD,\eD}(\e,n)\!=\!1]=1$.
\item
If $G$ is $\e$-far from being bipartite according to $\D$,
then $\prob[T^{G,\sD,\eD}(\e,n)\!=\!0]\geq2/3$. 
\EE
\ET
The Bipartite tester asserted in Theorem~\ref{bipartite:thm} 
is obtained by combining Theorem~\ref{bip-core:thm}
with a procedure that approximates the effective support size,
which in turn is provided in Section~\ref{approx-eff-support:sec}. 

\subsection{Proof of Theorem \protect\ref{bip-core:thm}}
Our starting point is the Bipartiteness tester of~\cite{GR2}
that works in the standard (bounded-degree graph) model.
On input an $n$-vertex graph, this tester operates by taking 
many (i.e., $\tildeO({\sqrt n})$) short (i.e., $\poly(\log n)$)
random walks from few (i.e., $O(1)$) randomly selected vertices
(where the $O$-notation hide a polynomial dependence on $1/\e$).  
In~\cite{GR2}, in each step of each random walk, 
the next vertex is selected uniformly among the neighbors
of the current vertex. Our basic idea is to adapt this tester
to the current setting by selecting the next vertex among 
the neighbors of the current vertex with probability that is 
{\em proportional to the probability weight of the corresponding 
incident edges}\/ (according to $\D$).

Wishing to use the analysis of~\cite{GR2} as a black-box,
we present the foregoing tester as emulating the tester of~\cite{GR2}
on an auxiliary graph in which weighted edges are replaced
by a proportional number of parallel edges, while recalling 
that {\em the analysis of~\cite{GR2} holds also for}\/ (non-simple) 
{\em graphs having}\/ (many) {\em parallel edges}\/ (see~\cite{KKR}). 
That is, on input $G$, we consider an auxiliary multi-graph $G'$ 
in which the number of edges between vertices is approximately
proportional to the weight of the corresponding edge in $G$. 
Indeed, this may yield a graph with a large discrepancy between
the average degree and the maximum degree,
but this can be fixed using the technique of~\cite{KKR}.
In fact, in order to present a simpler and more natural algorithm,
we use an alternative implementation of the foregoing technique, 
which suffices for our setting.

Fixing a vertex distribution $\D$, on input $G=(V,E)$, 
proximity parameter $\e$ and the value $n$
(provided as an effective support size of $\D$),
we consider the corresponding auxiliary multi-graph $G'=(V',E')$. 
For sake of simplicity, we shall assume first that 
{\em for each $v\in V$ either $\D(v)=0$ or $\D(v)\geq\rho$},
where we later set $\rho=\Theta(\e/n)$ 
and reduce the general case to this special case. 
%
Recalling that the weight of the edge $\{u,v\}\in E$ under $\D$
is $2\cdot(\D(u)+\D(v))/d$, 
we place $m_{u,v}\eqdef\round{(\D(u)+\D(v))\cdot N}\gg1$ 
parallel edges between $u$ and $v$ in $G'$, 
where $N=\poly(d/\e\rho)$ will be determined later. 
We keep only the non-isolated vertices in $V'$;
that is, $v\in V'$ if and only if $\sum_u m_{u,v}>0$,
which holds if and only if either $\D(v)\!\geq\!\rho$
or $\D(u)\!\geq\!\rho$ for some neighbor $u$ of $v$. 
Hence, $|V'|\leq d/\rho$. 

Indeed, the number of parallel edges between $u$ and $v$
is either zero or at least $\rho\cdot N-1$,
and the total the number of edges in $G'$ 
(i.e., $\sum_{\{u,v\}\in E}\round{(\D(u)+\D(v))\cdot N}$) 
is roughly $\sum_{w\in V}d_w\cdot\D(w)\cdot N=\Theta(N)$, 
where $d_v$ is the degree of $v$ in $G$. 
Hence, if $G$ is $\e$-far from being bipartite, under $\D$, 
then $G'$ is $(\e-o(\e))$-far from being bipartite 
(in the sense that more than $(\e-o(\e))\cdot|E'|$ edges
must be omitted from $G'$ to make it bipartite). 
Specifically, we have: 

\Bcm{\em(weight under $\D$ versus number of edges in $G'$):}
\label{G-vs-G':clm}
If edges of total $\D$-weight at least $\delta$ 
must be removed from $G$ in order to make it bipartite, 
then at least $\delta\cdot N-d|V'|$ edges must be removed from $G'$
in order to make it bipartite.  
\Ecm 
Indeed, if $N=\omega(|V'|/\delta)$,
then $\delta\cdot N-d|V'|=(\delta-o(\delta))\cdot N$. 
\medskip

\Bpf
Removing a weighted edge of $G$ is equivalent
to removing the corresponding parallel edges of $G'$.
Recalling that an edge $\{u,v\}\in E$ 
of weight $w_{u,v}=2\cdot(\D(u)+\D(v))/d\leq\D(u)+\D(v)$
yields $m_{u,v}=\round{(\D(u)+\D(v))\cdot N}\geq w_{u,v}\cdot N-1$ 
parallel edges and using $|E|<d|V'|$, the claim follows.%
\footnote{For any $E'\subseteq E$, we have 
\begin{eqnarray*}
\sum_{\{u,v\}\in E'} m_{u,v} 
&\geq& \sum_{\{u,v\}\in E'}(w_{u,v}\cdot N-1) \\ 
&=& \left(\sum_{\{u,v\}\in E'}w_{u,v}\right)\cdot N-|E'|. 
\end{eqnarray*}}
\Epf 
\medskip 

%

At this point, we could have emulated an execution
of the bipartite tester of~\cite{KKR} on $G'=(V',E')$,
which boils down to emulating the tester of~\cite{GR2}
on an auxiliary graph $G''=(V'',E'')$, where $|V''|=O(|V'|)$. 
(Such an emulation would have resulted in an algorithm
that is slightly different from the one we outlined above.) 
Instead, we prefer to capitalize on the specific features of $G'$,
and emulate the tester of~\cite{GR2} on a different auxiliary graph 
$G''=(V'',E'')$ that is more closely related to $G'$ (and to $G$). 
This allows us to invoke a minor variant of the tester of~\cite{GR2}
(indeed, the variant outlined upfront) directly on $G$. 
In particular, 
we shall use the fact that the number of parallel edges 
(between a pair of connected vertices) in $G'$ is large
(e.g., much larger than the square of the ratio between 
the maximum and average degrees in $G'$); that is, $M=\omega((N/M)^2/\e)$, 
where $M=\min_{\{u,v\}\in E:m_{u,v}>0}\{m_{u,v}\}\geq\rho\cdot N-1$. 
(Recall that the maximum degree in $G'$ is at most $N$,
whereas the average degree is at least $M$,
since each vertex in $G'$ has degree at least $M$.)

\paragraph{The auxiliary graph $G''=(V'',E'')$.}
As in~\cite{KKR}, we wish to transform $G'=(V',E')$ to 
a graph with a maximum degree that is only a constant factor
larger than its average degree, while preserving the
relative distance from the set of bipartite graphs. 
Denoting the degree of vertex $v$ (in $G'$)
by $d'_v=\sum_{u}m_{v,u}$ 
and letting $d'=\sum_{v\in V'}d'_v/|V'|=2|E'|/|V'|$
denote the average degree of $G'$, we replace each verterx in $G'$
by a cloud, denoted $C_v$, of $c_v\eqdef\ceil{d'_v/d'}$ vertices. 
Hence, $V''=\bigcup_{v\in V'}C_v$ 
and $|V''|=\sum_{v\in V'}|C_v|<\sum_{v\in V'}((d'_v/d')+1)=2|V'|$. 
(Unlike in~\cite{KKR}, we don't augment the graph with
bipartite expanders of degree $d'$ between $C_v$ and
an auxiliary set of $|C_v|$ vertices, but rather capitalize on 
the fact that the number of parallel edges between clouds
is much larger than the product of the size of the cluds;
that is, $m_{u,v}=\omega(|C_u|\cdot|C_v|/\e)$ 
for every $u,v\in V'$ such that $m_{u,v}>0$.)
Next, for each $u,v\in E'$ such that $m_{u,v}>0$, we partition 
the $m_{u,v}$ parallel edges that connect $u$ and $v$ (almost) equally
among the pairs of vertices in $C_u\times C_v$. 
That is, for every $\ang{u,i}\in C_u$ and $\ang{v,j}\in C_v$,
we connect $\ang{u,i}$ and $\ang{v,j}$ by $m_{u,v}^{i,j}$ parallel edges, 
where 
$m^{i,j}_{u,v}\in\{\floor{m_{u,v}/|C_u\times C_v|},
 \ceil{m_{u,v}/|C_u\times C_v|}\}$.

The key observation is that a random walk on $G'$
is closely related to a random walk on $G''$
in the sense that the walk on $G'$ moves from $u$ to $v$
with (almost) the same probability that the walk on $G''$
moves from a vertex in $C_u$ to a vertex in $C_v$.
This is the case since for every $i\in[c_u]$ it holds that 
$\frac{m_{u,v}}{\sum_{w}m_{u,w}}
 \approx\frac{m_{u,v}^{i,.}}{\sum_{w}m_{u,w}^{i,.}}$, 
where $m_{u,w}^{i,.}=\sum_{j\in[c_w]}m_{u,w}^{i,j}$. 
The approximation holds because $m^{i,.}_{u,w}\approx{m_{u,v}/c_u}$ 
which is due to $c_u \leq N/M \ll M$,
which in turn is due to $M\geq\rho N-1$ and $N=\omega(1/\rho^2)$.
Actually, to get an approximation factor of $1\pm\rho$,
we need $N/M \ll \rho\cdot M$, which holds when $N=\omega(1/\rho^3)$.
(Note that 
$\frac{m_{u,v}}{\sum_{w}m_{u,w}}
 \approx\frac{c_v\cdot m_{u,v}^{i,j}}{\sum_{w}m_{u,w}^{i,.}}$, 
holds if $(N/M)^2 \ll M$, but we don't use this fact.)

Indeed, $|E''|=|E'|$,
and so the average degree
(which is $2|E''|/|V''|<2|E''|/2|V'|=d'/2$)
is at least a third of the maximum degree 
(which is at most 
$\max_{\ang{v,j}\in V''}
 \{\sum_{\ang{u,i}\in V''}\ceil{m_{u,v}/c_uc_v}\}\leq d'+(d/\rho)<1.5d'$,
since $d'\geq M\geq \rho\cdot N-1=\omega(d/\rho)$).%
\footnote{To see the first inequality, 
note that $\sum_{\ang{u,i}\in V''}m_{u,v}/c_uc_v$
equals $\sum_{u\in V'}m_{u,v}/c_v=d'_v/c_v\leq d'$,
since $c_v=\ceil{d'_v/d'}\geq d'_v/d'$.
Hence, $\sum_{\ang{u,i}\in V''}\ceil{m_{u,v}/c_uc_v}$
is at most $d'+|V''|\leq d'+(d/\rho)$.}

We conclude that one can employ the tester of~\cite{GR2} to $G''$,
since the analysis in~\cite{GR2} presumes a constant
ratio between the maximum and the average degrees. 
Of course, we have to guaranteed that the distance
of $G$ to being bipartite is reflected by the distance
of $G''$ to being bipartite. This follows by combining
Claims~\ref{G-vs-G':clm} and~\ref{G'-vs-G'':clm}.

\Bcm{\em(number of violating edges in $G'$ versus $G'$):}
\label{G'-vs-G'':clm}
If at least $\Delta$ edges must be removed from $G'$
in order to make it bipartite, then
at least $\Delta-|V''|$ edges must be removed from $G''$
in order to make it bipartite.  
\Ecm 
The slackness is due to the fact that $m_{u,v}^{i,j}$
equals $\frac{m_{u,v}}{|C_u\times C_u|}$ up to at most one unit,
rather than being equal to it. Loosely speaking, this means that 
2-partitions of $G''$ with fewer inter-part edges are obtained 
by placing all vertices of each cloud on the same side. 
Intuitively, this holds because the edges between each pair of clouds
are partitioned equally between the corresponding pairs of vertices,
and so one does not benefit by treating vertices of the same cloud
differently (i.e., placing them on different sides of the 2-partition).
\medskip 

\Bpf
For every 2-partition $\chi'':V''\to\{1,2\}$,
we present a 2-partition $\chi':V'\to\{1,2\}$
such that the number of $\chi'$-monochromatic edges in $G'$
is at most $|V''|$ units larger than
the number of $\chi''$-monochromatic edges in $G''$;
that is, 
$$\sum_{u,v\in V':\chi'(u)=\chi'(v)}m_{u,v} \leq |V''| 
   +\sum_{\ang{u,i},\ang{v,j}\in V'':\chi''(\ang{u,i})=\chi''(\ang{v,j})}
       m_{u,v}^{i,j}.$$
This is shown by using the probabilistic method. 
Specifically, fixing $\chi'':V''\to\{1,2\}$,
we consider assigning each vertex of $G'$
a color chosen at random according to the colors of the
vertices in the corresponding cloud; that is, $\chi'(v)=1$
with probability $|\{i\in[c_v]:\chi''(\ang{v,i})=1\}|/|C_v|$, 
and $\chi'(v)=2$ otherwise. 
Letting $X_{u,v}$ be a random variable indicating
whether or not $\chi'(u)=\chi'(v)$  
(i.e., $X_{u,v}=1$ if $\chi'(u)=\chi'(v)$ and $X_{u,v}=0$ otherwise), 
we have 
\begin{eqnarray*}
\Exp\left[\sum_{u,v\in V'}m_{u,v}\cdot X_{u,v}\right]
&=& \sum_{u,v\in V'}m_{u,v}\cdot\prob[\chi'(u)=\chi'(v)] \\ 
&=& \sum_{u,v\in V'}\frac{m_{u,v}}{|C_u\times C_u|}
    \cdot|\{(i,j)\in[c_u]\times[c_v]:\chi''(\ang{u,i})=\chi''(\ang{v,j})\}|. 
\end{eqnarray*}
Recalling that the absolute difference between $m_{u,v}^{i,j}$
and $\frac{m_{u,v}}{|C_u\times C_u|}$ is at most one unit,
it follows that the expected number of $\chi'$-monochromatic edges in $G'$
differs from the number of $\chi''$-monochromatic edges in $G''$
by at most $|V''|$ units. 
\Epf 

\paragraph{The actual tester.}
Recall that, when given oracle access to an $n$-vertex graph
(and proximity parameter $\e$),
the tester of~\cite{GR2} selects uniformly $O(1/\e)$ (start) vertices,
and starts $\poly(1/\e)\cdot\tildeO({\sqrt n})$
random $\ell$-step walks from each vertex, 
where $\ell=\poly(\e^{-1}\log n)$
and each step in the random walk moves uniformly
to one of the neighbors of the current vertex.
The tester accepts if and only if the explored subgraph 
is bipartite (i.e., contains no odd-length cycles). 
When seeking to test $G$ under the vertex-distribution $\D$,
we shall emulate a testing of $G''$ as follows.
\BI
\item The $O(1/\e)$ start vertices will be selected
according to the degrees in $G'$ (and in $G''$), 
which in turn reflect the weight of the incident edges. 
Hence, we wish to select $v$ with probability that is 
proportional to $\sum_{u\in\Gamma(v)}(\D(u)+\D(v))$,
where $\Gamma(v)$ denotes the set of neighbors of $v$ in $G$. 
We do so by employing ``rejection sampling'' as follows.
First, we obtain a sample of $\D$; that is, $s\gets\D$. 
Next, we output $s$ with probability $|\Gamma(s)|/2d$,
and halt with no output with probability $(2d-2|\Gamma(s)|)/2d$,
Otherwise (i.e., with probability $|\Gamma(s)|/2d$),
we output a uniformly selected neighbor of $s$
(i.e., each neighbor of $s$ is output with probability $1/2d$). 

Hence, the probability that $v$ is output (in a single trial) equals  
$p(v)\eqdef\D(v)\cdot\frac{|\Gamma(v)|}{2d}
 +\sum_{u\in\Gamma(v)}\D(u)\cdot\frac1{2d}$,
where the first term is due to selecting $s=v$
and the second term is due to selecting $s\in\Gamma(v)$. 
Note that $p(v)=\sum_{u\in\Gamma(v)}(\D(v)+\D(u))/{2d}$,
whereas the degree of $v$ in $G'$ is proportional to 
$\sum_{u\in\Gamma(v)}m_{u,v}\approx\sum_{u\in\Gamma(v)}(\D(u)+\D(v))\cdot N$. 
(Indeed, we repeat trying till some vertex is output,
while noting that each trial succeeds at least with probability $1/d$.)

\item A step in a random walk is made by selecting
a neighbor of the current vertex with probability
that is proportional to the weight of the edge leading to it.
Specifically, when we reside at $v\in V$,
we move to the neighbor $u$ with probability
proportional to $\D(v)+\D(u)$. 
\EI 
Note that for implementing the foregoing actions 
it suffices to be able to sample $\D$, 
and have oracle access to the evaluation of $\D$
(and to the incidence function of $G$). 
Actually, the evaluator of $\D$ can be replaced
by an oracle that answers $(w_1,w_2)$ with $\D(w_1)/\D(w_2)$,
provided $\D(w_2)>0$ (and a special symbol otherwise).

\paragraph{The general case.}
Recall that the foregoing analysis presumes that 
for each $v\in V$ either $\D(v)=0$ or $\D(v)\geq\rho$.
Using $\rho=\Theta(\e/n)$, we now reduce the general case 
of $\D$ that has $\e/4$-effective support size $n$ 
to the foregoing case.
We first note that, by the following Claim~\ref{effective:clm},
the vertex-distribution $\D$ is $\e/2$-close
to a vertex-distribution $\D'$ that satisfiers
the foregoing condition with $\rho=\e/4n$. 
Hence, if $G$ is $\e$-far from being bipartite under $\D$,
then it is $0.5\e$-far from being bipartite under $\D'$,
and we can test $G$ by providing the tester 
with oracle access to $\D'$ (i.e., to $\sD'$ and $\eD'$) 
and setting the proximity parameter to $\e/2$.
Actually, it suffices to provide the foregoing tester
with sampling access to $\D'$, 
which can be emulated by ``rejection sampling'' via $\D$,
and with an evaluator of the ratios between the $\D'$-values 
(which, in turn, equal the corresponding the ratios between the $\D$-values). 

\Bcm{\em(effective support size and minimal weight):}
\label{effective:clm}
Suppose that $\D$ has an $\eta$-effective support of size $n$. 
Then, $\D$ is $2\eta$-close to a distribution $\D'$ 
that satisfies $\min_{e:\D'(e)>0}\{\D'(e)\}>\eta/n$. 
Furthermore, $\D'(e)>0$ if and only if $\D(e)>\eta/n$,
and if $\D'(e)>0$ then $\frac{\D'(e')}{\D'(e)}=\frac{\D(e')}{\D(e)}$
for every $e'$. 
\Ecm
Combining the foregoing analysis with Claim~\ref{effective:clm},
Theorem~\ref{bip-core:thm} follows. 
\medskip

\Bpf
We may assume, without loss of generality, 
that the support of $\D$ has $s>n$ elements, 
and arrange these elements according to their $\D$-value;
specifically, let $e_1,...,e_s$ such that $\D(e_i)\geq\D(e_{i+1})>0$
for every $i\in[s-1]$. Then, $\sum_{i=n+1}^s\D(e_i)\leq\eta$,
because the distance of $\D$ from the class of distributions
of support size at most $n$ is at least $\sum_{i=n+1}^s\D(e_i)$.
Letting $P(e)=\D(e)$ if $\D(e)>\eta/n$ and $P(e)=0$,
we define $\D'(e)=\frac{P(e)}{\sum_{i\in[s]}P(e_i)}$,
and notice that 
\begin{eqnarray*}
\sum_{e:P(e)=0}\D(e)
&\leq& \sum_{i\in[n]:\D(e_i)\leq\eta/n}\D(e_i)
   + \sum_{i\in[s]\setminus[n]}\D(e_i) \\
&\leq& 2\eta. 
\end{eqnarray*}
Noting that $\D'(e)\geq\D(e)$ for every $e\not\in P^{-1}(0)$,
the main claim follows. The furthermore claim follows
by the definition of $\D'$. 
\Epf 

%

\subsection{Approximating the effective support size of $\D$}
\label{approx-eff-support:sec}
The foregoing Bipartite tester,
which establishes Theorem~\ref{bip-core:thm},
presupposes that the tester is given an upper bound 
on the minimum $\e/4$-effective support size of $\D$ as auxiliary input. 
Proving Theorem~\ref{bipartite:thm} requires
getting rid of that auxiliary input, or rather approximating it
when using oracle access to $\D$. 
Indeed, we shall show that given sampling and evaluation oracles to $\D$,
it is relatively easy to approximate its effective support size.
(In contrast, as commented in the introduction, 
obtaining such an approximation while using only samples of $\D$ 
is too expensive for our application.)

Note that the notion of approximating the effective support size 
is not robust in the sense that, for every $\eta<\eta'$,
a distribution $\D$ may be have a minimal $\eta$-effective support size 
that is much larger than its minimal $\eta'$-effective support size
(e.g., consider $\D$ that assigns a total probability mass
of~$1-\eta$ to very few elements and is otherwise uniform
on a huge set). 
Hence, on input $\eta$, our goal is to approximate 
the $\Theta(\eta)$-effective support size of $\D$; say, output 
a number between the minimal $2\eta$-effective support size of $\D$
and its minimal $\eta$-effective support size 
(or a good approximation of such a number). 

\paragraph{A simple approximation of the effective support size.}
We first present a simple algorithm for obtaining a rather
rough (but sufficiently good for our purposes) approximation. 
Given an ``effectiveness'' parameter $\eta$, 
we proceed in iterations such that in the $i^\xth$ iteration 
we {\em take a sample of $m=O(\eta^{-1}\log i)$ elements of $\D$,
and halt outputting $2^i/\eta$ if the number of samples 
that have $\D$-value below $\eta/2^i$ is at most $3\eta\cdot m$}. 
Observer that (in iteration $i$), 
with probability at least $1-0.01/i^2$, 
the sample approximates the total weight of the ``light elements''
(i.e., elements having $\D$-value below $\eta/2^i$)
in the sense that
if $\prob_{v\gets\D}[\D(v)<\eta/2^{i}]<2\eta$,
(resp., if $\prob_{v\gets\D}[\D(v)<\eta/2^{i}]\geq4\eta$),
then the number of samples that have $\D$-value below $\eta/2^i$ 
is at most $3\eta\cdot m$ (resp., is greater than $3\eta\cdot m$). 

Now, letting $n$ be an upper bound on the $\eta$-effective support size 
of $\D$ (and assuming for simplicity that $n$ is a power of two),
observe that, with high (constant) probability,
we halt by iteration $i^*=\log_2n$, 
because $\prob_{v\gets\D}[\D(v)<\eta/2^{i^*}]<n\cdot\eta/2^{i^*}+\eta=2\eta$,
where the first (resp., second) term is due to elements 
that are (resp., are not) in the effective support of $\D$. 
Hence, with high (constant) probability,
the algorithm outputs a value that is at most $n/\eta$.   
(Actually, if we reach iteration $i^*$, we halt in it with
probability at least $1-0.01/(i^*)^2$; it follows that
with probability at $1-0.01/\log^2 n$, 
the algorithm outputs a value that is at most $n/\eta$.)   

On the other hand, assuming that the minimal $4\eta$-effective 
support size of $\D$ is at least $n'$
(and assuming that $n'$ is also a power of two), 
with high (constant) probability,
we do not halt before iteration $i^+=\log_2(\eta\cdot n')$, 
because otherwise $\prob_{v\gets\D}[\D(v)<\eta/2^{i^+-1}]<4\eta$, 
which implies that $\D$ has $4\eta$-effective support size at most $n'/2$ 
(since $\prob_{v\gets\D}[\D(v)<2/n']<4\eta$ implies that $\D$ 
is $4\eta$-close to a distribution of support size at most $n'/2$).
Hence, with high (constant) probability
(i.e., with probability at least $1-\sum_{i<i^+}0.01/i^2>0.98$), 
the algorithm outputs a value that is at least $n'$. 

It follows that, with high (constant) probability,
the algorithm outputs a number that lies between $n'$ and $n/\eta$
(i.e., between the half the minimal $4\eta$-effective support size of $\D$
and $2/\eta$ times its minimal $\eta$-effective support size). 
Furthermore, with high (constant) probability, 
this algorithm runs for 
$\sum_{i\leq i^*} O(\eta^{-1}\cdot\log i) = \tildeO(\log n)/\eta$  
steps (and its expected number of steps can be similarly bounded).
Combining this result with Theorem~\ref{bip-core:thm},
we essentially infer Theorem~\ref{bipartite:thm}
(except that $\e/5$ is replaced by $\e/16$). 

\paragraph{Obtaining better approximations of the effective support size.}
For starters, we present a tighter analysis of (a minor variant of)
the foregoing algorithm. Specifically, for any constant $\beta>1$,
in the $i^\xth$ iteration, we {\em take a sample of $m=O(\eta^{-1}\log i)$ 
elements of $\D$, and halt outputting $2^i/\eta$ if the number of samples 
that have $\D$-value below $(\beta-1)\cdot\eta/2^i$ is 
at most $\beta^2\cdot\eta\cdot m$}. 
In analyzing the probability that this algorithm halts 
by iteration $i^*=\log_2n$, 
we use the fact that
$\prob_{v\gets\D}[\D(v)<(\beta-1)\cdot\eta/2^{i^*}]
 <n\cdot(\beta-1)\cdot\eta/2^{i^*}+\eta=\beta\cdot\eta$,
whereas in analyzing the probability that the algorithm 
does halts before iteration $i^+=\log_2(\eta\cdot n')$ implies 
$\prob_{v\gets\D}[\D(v)<(\beta-1)\cdot\eta/2^{i^+-1}]<\beta^3\cdot\eta$
(where here $n'$ is the minimal $\beta^3\eta$-effective support size).%
\footnote{Here we use the fact that (in iteration $i$), 
with probability at least $1-0.01/i^2$, 
the sample approximates the total weight of the ``light elements''
(i.e., elements having $\D$-value below $\eta/2^i$)
in the sense that
if $\prob_{v\gets\D}[\D(v)<(\beta-1)\cdot\eta/2^{i}]<\beta\cdot\eta$,
(resp., if 
$\prob_{v\gets\D}[\D(v)<(\beta-1)\cdot\eta/2^{i}]\geq\beta^3\cdot\eta$),
then the number of samples that have $\D$-value below $(\beta-1)\cdot\eta/2^i$ 
is at most $\beta^2\cdot\eta\cdot m$ 
(resp., is greater than $\beta^2\cdot\eta\cdot m$).}
It follows that, with high (constant) probability,
the algorithm outputs a number, denoted $\nn$, that lies between 
half the minimal $\beta^3\cdot\eta$-effective support size of $\D$
and $2/\eta$ times its minimal $\eta$-effective support size. 

To obtain an even better approximation of the effective support size,
we use the rough estimate $\nn$ obtained above
in order to approximate the number of elements
that have $\D$-value approximately $\beta^{i-0.5}$
for every $i\in[O(\log\nn/\eta)]$. 
Indeed, our first step is ignoring elements having $\D$-value
below $\eta/\nn$ or so. 
Specifically, setting $\eta'=\beta^3\cdot\eta$, 
recall that if $\D$ has an $\eta'$-effective support of size $\nn$, 
then $\D(H)\geq1-\beta\eta'$ 
for $H\eqdef\{v:\D(v)\geq(\beta-1)\cdot\eta'/\nn\}$ 
(see prior paragraph, while replacing $i^*$ by $\log_2\nn$).%
\footnote{In other words,
observe that 
$\prob_{v\gets\D}[\D(v)<(\beta-1)\cdot\eta'/\nn]
 <\nn\cdot(\beta-1)\cdot\eta'/\nn+\eta'=\beta\cdot\eta'$,
where the first (resp., second) term is due to elements 
that are (resp., are not) in the effective support of $\D$.}
Hence, assuming that $\D(H)\leq1-\eta$ 
and letting $\eta''=\beta\cdot\eta'=\beta^4\cdot\eta$, 
it holds that $|H|$ lies 
between the minimal $\eta''$-effective support size of $\D$
and its minimal $\eta$-effective support size, and so 
providing a good approximation of the ``effective size'' of $H$ will do. 
(In the case that $\D(H)>1-\eta$ additional steps will be needed.) 

To (effectively) approximate $|H|$, 
we let $W_i=\{v:\beta^{i-1}\leq\D(v)<\beta^i\}$, 
and observe that it suffice the approximate $\D(W_i)$ for $i=1,...,\ell$, 
where 
$\ell\eqdef\log_\beta(\nn/(\beta-1)\cdot\eta')
 =O((\beta-1)^{-1}\cdot\log(\nn/\eta))$. 
Actually, letting $I=\{i\in[\ell]:\D(W_i)\geq(\beta-1)\eta''/\ell\}$,
it suffices to approximate $\D(W_i)$ for every $i\in I$,
which yields approximations of the corresponding $|W_i|$'s
(using $|W_i|\approx\D(W_i)/\beta^{i-0.5}$). 
That is, we do not actaully approximate $|H|$ but
rather approximate the size of $H'\eqdef\bigcup_{i\in I}W_i\subseteq H$,
while capitalizing on $\D(H\setminus H')\leq(\beta-1)\cdot\eta''$. 
Hence, we will approximate the minimal $\eta'''$-support size
for some $\eta\leq\eta'''\leq\beta\eta''=\beta^5\eta$. 
Specifically, for each $i\in[\ell]$, 
using a sample of $O(t\ell/(\beta-1)^{2}\cdot\eta')$ samples,
we obtain (with probability $1-2^{-t}$)
a $\beta$-factor approximation of $\D(W_i)$ for each $i\in I$,
which yields a $\beta^2$-factor approximation of $|W_i|$
(since $|W_i|\in[\beta^{i-1}\D(W_i),\beta^i\cdot\D(W_i))$). 
Note that the rough estimate of the effective support size of $\D$
(i.e., $\nn$) is only used in order to determine $\ell$. 

Recall that we have assumed that $\D(H)\leq1-\eta$,
whereas this is not necessarily the case.
Needless to say, we can easily estimate $1-\D(H)$ up to any 
desired constant factor (using $O(1/\eta)$ samples of $\D$). 
In case we are quite sure that $\D(H) > 1-\eta$
(which will happen if $\D(H) > 1-\beta\eta$
but not if $\D(H) < 1-\beta^{-1}\eta$), 
we can reduce the estimate obtained for $|H|$ by disposing 
an adequate weight of $H$;
that is, we dispose as many as the the lightest elements 
as possible till reaching a subset of $H'$ that has $\D$-value 
that is most likely to be below $1-\eta$. 
Note that the foregoing process is performed
without making any samples or queries; it is solely based on 
the estimated values of $\D(W_i)$ for $i\in I$ already obtained. 
Hence, we get.

\BT{\em(approximating the effective support size):} 
\label{approx-eff:thm}
There exists an oracle machine $M$ that, 
on input $\eta$ and $\beta>1$, 
satisfies the following condition for every distribution $\D$. 
Given oracle access to $\sD$ and $\eD$,
with probability at least $2/3$, the machine outputs a number $n$ 
after making $\poly(1/(1-\beta))\cdot\tildeO(\eta^{-1}\log n)$ steps
such that $n$ is at least the minimal $\beta^5\cdot\eta$-effective support
size of $\D$ and is at most $\beta^2$ times
the minimal $\beta^{-1}\cdot\eta$-effective support size of $\D$. 
Furthermore, the expected number of steps performed by $M$
is $\poly(1/(1-\beta))\cdot\tildeO(\eta^{-1}\log n)$.
\ET
Needless to say, by a change in parameters we can make $n$ lie
between the minimal $\beta\cdot\eta$-effective support size of $\D$ 
and $\beta$ times its minimal $\eta$-effective support size.
Combining Theorems~\ref{bip-core:thm} and~\ref{approx-eff:thm},
we infer Theorem~\ref{bipartite:thm}. 

The approximation algorithm of Theorem~\ref{approx-eff:thm}
provides a rather good approximation of the effective support size,
but its complexity depends (mildly) on the effective support size.
We wonder whether it is possible to obtain a meaningful approximation
of the effective support size within complexity that is independent of it. 
For starters, can approximations as in Theorem~\ref{approx-eff:thm}
be obtained within complexity $\poly(1/\eta)$? 
\ifnum\eess=1 
In a forthcoming paper, 
we farther explore the complexity of approximating the
effective support size of distributions. 
We present several algorithms that exhibit a trade-off between 
the quality of the approximation and the complexity of obtaining it,
and leave open the question of their optimality.

\else 
\BOP{\em(faster approximating the effective support size):} 
\label{approx-eff:open}
Do there exist a function $T:\N\to\N$ and an oracle machine that, 
on input $\eta>0$ and oracle access to $\sD$ and $\eD$,
runs for expected $T(1/\eta)$-time and outputs, 
with probability at least $2/3$, a number that lies
between the minimal $O(\eta^{\Omega(1)})$-effective support size of $\D$ 
and $T(1/\eta)$ times its minimal $\eta$-effective support size?
\EOP
Of course, $f(\eta)=O(\eta^{\Omega(1)})$ may be replaced by 
any function $f:[0,1]\to[0,1]$ that tends to zero
(i.e., for every $\e>0$ there exists $\eta>0$
such that $f(\eta)\leq\e$). 
\fi 

\ifnum\noTitle=1 
\paragraph{Approximating the effective support size in a weaker model.}
Here we consider a model in which we cannot evaluate $\D$ at given points,
but can obtain the ratio of $\D$-values between any two points.
Note that in this model distinguishing 
a uniform distribution on $n$ elements from 
a uniform distribution on $n/2$ elements 
requires $\Omega({\sqrt n})$ queries.
We shall show that this lower bound can be met.

The basic idea is to take samples of $\D$ until a sufficiently high 
constant number of pairwise collisions (among the samples) are seen. 
(These pairwise collisions may represent higher order collisions
as well as disjoint pairwise collisions.)
If some sample appears sufficiently many times,
we may afford to approximate its $\D$-value (by additional sampling),
and using this value obtain the $\D$-value of 
any other element by comparison. 
Otherwise, we fix an arbitrary point~$p$ that appears in the sample, 
tentatively set its $\D$-value to equal the reciprocal of
the square of the size of the sample, 
and use this value to tentatively set the $\D$-value 
of any other element by comparison. 
In both cases, we use the same procedure as in the case
that we do have access to an evaluator of $\D$,
but in the second case we first use the procedure 
to compute the ``tentative probability'' that $\D$ assigns its support
and normalize the $\D$-value of~$p$ accordingly. 

Note that in both cases, the size of the initial sample
is at most a square root of the size of the support,
since the collision probability of a distribution
with support size $n$ is at least $1/n$.  
\fi 

\section{The Cycle-freeness Tester}\label{cycle:sec}
Following~\cite{CGRSSS}, we reduce testing cycle-freeness
to testing bipartiteness, where in both cases we refer
to VDF testing in the bounded degree graph model. 
While the reduction is identical to the one presented in~\cite{CGRSSS} 
(for the standard bounded-degree model), 
our analysis presented in Lemma~\ref{cycle-free2bipartite:lem}
addresses issues that arise only in the VDF model. 

Specifically, we use the presentation provided in~\cite[Sec.~8.1]{CGRSSS} 
rather than the one provided in~\cite[Sec.~3]{CGRSSS}.
The pivot of this presentation is the following 
{\em generalization of 2-colorability}\/
in which edges of the graph are labeled
by either $\tt eq$ or $\tt neq$. That is, an instance
of this problem is a graph $G=(V,E)$ along with
a labeling $\tau:E\to\{{\tt eq},{\tt neq}\}$.
We say that $\chi:V\to\bitset$ is a {\sf legal 2-coloring} 
of this instance if for every $\{u,v\}\in E$
it holds that $\chi(u)=\chi(v)$ if and only if $\tau(\{u,v\})={\tt eq}$.
That is, a legal 2-coloring (of the vertices) is one in which
every two vertices that are connected by an edge labeled $\tt eq$
(resp., $\tt neq$) are assigned the same color (resp., opposite colors).

Note that the standard notion of 2-colorability corresponds
to the case in which all edges are labeled $\tt neq$.
We first observe that the Bipartite tester provided
in Section~\ref{bip:sec}
can be extended to test this generalization of 2-colorability.
This is the case because 
the (one-sided error) Bipartite tester of~\cite{GR2} 
(as well as the ones of~\cite{KKR} and~\cite{GGR})
can be extended to test this generalization of 2-colorability.

Next, as in~\cite[Sec.~8.1]{CGRSSS}, we randomly reduce testing 
cycle-freeness of a graph $G=(V,E)$ to testing generalized 2-coloring
of the instance obtained by coupling $G$ with a uniformly selected
labeling $\tau:E\to\{{\tt eq},{\tt neq}\}$,
which may be selected on the fly
(i.e., whenever we encounter a new edge, we assign it a random label).
Clearly, if $G$ is cycle-free,
then, for any choice of $\tau:E\to\{{\tt eq},{\tt neq}\}$,
the instance $(G,\tau)$ has a legal 2-coloring. 
We conjecture that, like in~\cite[Lem.~3.1]{CGRSSS}, it holds that
if $G$ is $\e$-far from being cycle-free (under the distribution $\D$),
then in expectation the random instance $(G,\tau)$ 
is $\Omega(\e)$-far from having a legal 2-coloring 
(under the distribution $\D$).
We only prove a weaker result, which suffices for our application
(since $\tildeO({\sqrt n})\cdot\poly(\Omega(\e/\log n)^{-1})
  = \tildeO({\sqrt n})\cdot\poly(\e^{-1})$).

\BL{\em(analysis of the randomized reduction):}
\label{cycle-free2bipartite:lem}
Let $G=(V,E)$ be a simple graph
{\rm(i.e., $G$ has neither parallel edges nor self-loops)}%
\footnote{Alternatively, we consider a pair parallel edges 
(resp., a self-loop) as constituting a cycle.} 
and $\D$ be a distribution on $V$. 
If $G$ is $\e$-far from being cycle-free according to distribution $\D$, 
then, with probability at least $\Omega(1)$ 
over the random choice of $\tau:E\to\{{\tt eq},{\tt neq}\}$,
the instance $(G,\tau)$ is $\Omega(\e/\log|V|)$-far from 
having a 2-legal coloring.
\EL
Hence, we reduce testing cycle-freeness of $n$-vertex graphs
with respect to the proximity parameter $\e$ and vertex-distribution $\D$
to testing generalized 2-coloring of $n$-vertex graphs
with respect to the proximity parameter $\Omega(\e/\log n)$
(and vertex-distribution $\D$). 
An analogous assertion holds for testing cycle-freeneess
with respect to arbitrary graphs and distribution $\D$
that have $\e/5$-effective support size $n$. 
Hence, combining Lemma~\ref{cycle-free2bipartite:lem}
and Theorem~\ref{bipartite:thm},
we obtain Theorem~\ref{cycle-free:thm}. 
\medskip 


\BPF
Maintaining $\D$ unchanged,
we first omit from $G$ all edges that have probability weight 
smaller than $\e/d|V|$ (i.e., we omit the edge $\{u,v\}$
if $2\cdot\frac{\D(u)+\D(v)}{d}<\e/d|V|$). 
Denoting the resulting graph by $G'=(V,E')$, observe that $G'$ 
is $\e/2$-far from being cycle-free according to distribution $\D$.  
Next, we consider a spanning forest of maximal weight of $G'$,
denote its edges by $F'$, and note that the weight of
the edges in $E'\setminus F'$ is at least $\e/4$,
because $G'$ can be made cycle-free by omitting the edges $E'\setminus F'$. 
We partition the edges $E'\setminus F'$ to buckets according to their weight,
letting the bucket $B_i\subseteq E'\setminus F'$ contain the edges of weight 
in $(2^{-i},2^{-i+1}]$. 
Letting $\ell=\floor{\log_2(d|V|/\e)}+1$,
we observe that $E'\setminus F'=\bigcup_{i\in[\ell]}B_i$,
and it follows that there exists $i\in[\ell]$ such that 
the total weight of the edges in $B_i$ is at least $\e/4\ell$. 

Fixing such an $i\in[\ell]$ and 
focusing on the graph $G''=(V,F'\cup B_i)$,
we consider the two-connected components of $G''$ 
(i.e., the maximal subgraphs of $G''$ in which every two
vertices are connected by at least two edge-disjoint paths). 
We observe that {\em edges that connect such components must be in $F'$},
by considering two cases. 
\begin{quote}
In the first case, these components are not connected in $G'$, 
and in this case there are also not connected in $G''$ 
(and the claim holds vacuously). 
Otherwise (which is the second case), these components
are connected in $G'$ and so there must be a path of edges of $F'$
that connects (a pair of vertices in) them, 
since $F'$ spans the connected components of $G'$. 
In fact, in this case, any pair of vertices in these two components 
is connected by a path of edges in $F'$. 
But, then having an edge of $B_i$ that connects 
these components yields a contradiction to their definition 
(as two-connected components of $G''$), since it yields two
enge-disjoint paths between the endpoints of this edge
(one being the edge itself and the other being a path of edges in $F'$).
\end{quote}
Lastly, we omit the edges of $F'$ that connect different
two-connected components, and obtain a graph $G'''$
in which each connected component is two-connected. 

We next claim that each edge in $G'''=(V''',E''')$ has weight
at least $2^{-i}$. Since $E'''\subseteq F'\cup B_i$,
we have to establish this claim only for edges in $E'''\cap F'$. 
We first observe that the edges in $E'''\cap F'$ induce 
a spanning tree of each connected component of $G'''$,
because otherwise we reach a contradiction to their definition
(as two-connected components of $G''$).%
\footnote{Specifically, vertices in a component of $G'''$
that are not connected by edges of $F'$ must be connected in $G''$ 
by an edge-disjoint path of edges of $F'$ that contain vertices
that are not in this component. But this contradicts the hypothesis
that this (two-connected) component of $G'''$ is a two-connected
component of $G''$ (i.e., it contradicts its maximality in $G''$).}
Suppose towards the contradiction that one of the edges
of such a spanning tree has weight smaller than $2^{-i}$.
Then, omitting this edge and adding an edge of $B_i$
that connects the two resulting sub-trees,
where such an edge must exist by two-connectivity of the component,
we obtain a spanning tree of this component that has
a larger weight than the original one. 
This yields a spanning forest of $G'$ that has weight
larger than the forest $F'$, 
in contradiction to the definition of $F'$. 

Finally, we apply~\cite[Lem.~3.1]{CGRSSS} to $G'''$,
while noting that $E'''\setminus F'=B_i$
and $|B_i|\geq\frac{\e/4\ell}{2^{-i+1}}=\Omega(2^i\cdot\e/\ell)$,  
it follows that 
(with probability at least $\Omega(1)$ over the choice of $\tau$)
at least $\Omega(|B_i|)$ edges must be omitted from $G'''$
such that the resulting instance $(\cdot,\tau)$ has a legal 2-coloring.
Recalling that each edge in $G'''$ have weight at least $2^{-i}$,
it follows that the weight of edges that must be omitted from $G'''$
in order to yield an instance $(\cdot,\tau)$ that has a legal 2-coloring
is at least $\Omega(|B_i|\cdot2^{-i})=\Omega(\e/\ell)$.  
The same holds with respect to $G$, since $G'''$ is a subgraph of $G$,
and the claim follows. 
\EPF

\section*{Acknowledgments}
This project was partially supported by 
the Israel Science Foundation (grant No.~1146/18), 
and has received funding from the European Research Council (ERC) 
under the European Union's Horizon 2020 research and innovation programme 
(grant agreement No. 819702).


\addcontentsline{toc}{section}{Bibliography}


\begin{thebibliography}{ABCD}

\bibitem{BCG} E.~Blais, C.L.~Canonne, and T.~Gur.
\newblock Distribution Testing Lower Bounds 
          via Reductions from Communication Complexity. 
\newblock In {\em 32nd Computational Complexity Conference}, 
          pages 28:1--28:40, 2017. 

\bibitem{CGRSSS} 
A.~Czumaj, O.~Goldreich, D.~Ron, C.~Seshadhri, A.~Shapira, and C.~Sohler. 
\newblock Finding cycles and trees in sublinear time.
\newblock {\em RS\&A}, Vol.~45(2), pages 139--184, 2014. 

\bibitem{G:pt} O.~Goldreich.
\newblock {\em Introduction to Property Testing}. 
\newblock Cambridge University Press, 2017.
\bibitem{G:vdf} O.~Goldreich.
\newblock Testing Graphs in Vertex-Distribution-Free Models. 
\newblock {\em ECCC}, TR18-171, 2018. 
\newblock (See Revision Nr~1, March 2019.)

\bibitem{GGR} O.~Goldreich, S.~Goldwasser, and D.~Ron.
\newblock Property testing and its connection to learning and approximation.
\newblock {\em Journal of the ACM}, pages 653--750, July 1998.

\bibitem{GR} O.~Goldreich and D.~Ron.
\newblock Property Testing in Bounded Degree Graphs. 
\newblock {\em Algorithmica}, Vol.~32~(2), pages 302--343, 2002.
\bibitem{GR2} O.~Goldreich and D.~Ron.
\newblock A Sublinear Bipartitness Tester for Bounded Degree Graphs.
\newblock {\em Combinatorica}, Vol. 19~(3), pages 335--373, 1999.

\bibitem{KKR} T.~Kaufman, M.~Krivelevich, and D.~Ron.
\newblock Tight Bounds for Testing Bipartiteness in General Graphs.
\newblock {\em SIAM Journal on Computing}, Vol.~33~(6), pages 1441--1483, 2004.

\bibitem{RRSS} S.~Raskhodnikova, D.~Ron, A.~Shpilka, and A.~Smith. 
\newblock Strong Lower Bounds for Approximating Distribution Support Size 
          and the Distinct Elements Problem. 
\newblock {\em SICOMP}, Vol.~39~(3), pages 813--842, 2009. 

\end{thebibliography}
\end{document}